\documentstyle[12pt]{article}
\begin{document}
\newcommand{\newc}{\newcommand}
\newc{\gev}{\,GeV}
\newc{\rp}{$R_p$}
\newc{\rpv}{$\not\!\!R_p$}
\newc{\rpvm}{{\not\!\! R_p}}
\newc{\rpvs}{{\not R_p}}
\newc{\ra}{\rightarrow}
\newc{\lra}{\leftrightarrow}
\newc{\lsim}{\buildrel{<}\over{\sim}}
\newc{\gsim}{\buildrel{>}\over{\sim}}
\newc{\esim}{\buildrel{\sim}\over{-}}
\newc{\lam}{\lambda}
\newc{\lsp}{{\tilde\Lambda}}
\newc{\oc}{{\cal {O}}}
\newc{\msq}{m_{\tilde q}}
\newc{\mpl}{M_{Pl}}
\newc{\mw}{M_W}
\newc{\pho}{{\tilde\gamma}}
\newc{\half}{\frac{1}{2}}
\newc{\third}{\frac{1}{3}}
\newc{\quarter}{\frac{1}{4}}
\newc{\beq}{\begin{equation}}
\newc{\eeq}{\end{equation}}
\newc{\barr}{\begin{eqnarray}}
\newc{\earr}{\end{eqnarray}}
\newc{\delb}{\Delta B\not=0}
\newc{\dell}{\Delta L\not=0}
\newc{\delbi}{\Delta B_i\not=0}
\newc{\delli}{\Delta L_i\not=0}
\title{ Sphaleron Erasure of Primordial Baryogenesis}
\author{H.^Dreiner and G.^G.^Ross\thanks{SERC Senior Fellow}}
\date{{\small Department of Physics, University of Oxford,\\
  1 Keble Rd, Oxford OX1 3NP}}
\maketitle

\begin{abstract}
\noindent If the present baryon-asymmetry is due to a Planck or GUT-scale
matter
asymmetry then  baryon- or lepton-number violating processes are constrained by
the condition that they do not subsequently erase this asymmetry. We present a
revision of the analysis of sphaleron baryon-number violating processes in the
standard model including lepton-mass effects. We find the surprising result
that
a GUT-scale matter-asymmetry can survive the $B$ and $L$ violating sphaleron
interactions even though ($B- L$) is conserved and equals zero for all
temperatures. We extend the analysis to cover the minimal supersymmetric
standard model (MSSM) and also derive the constraints on the R-parity violating
couplings in extensions of the MSSM. In the case of the baryon number violating
dimension 4 operators we find, contrary to current wisdom, that the resulting
bounds can be avoided completely because of a residual lepton-flavour number
conservation; in the case of lepton number violating operators we find the
bounds are flavour dependent and can be avoided completely in definite flavour
channels. We also consider how the bounds are modified in the case there is a
Grand Unified extension of the supersymmetric model which introduces new lepton
flavour violating couplings.
\end{abstract}

\section{Introduction}
It has recently been observed  \cite{campbell}-\cite{harvey} that the
requirement of a cosmological baryon-asymmetry leads to powerful constraints on
the initial nature of the asymmetry and on extensions of the Standard Model
(SM)
which introduce new baryon- or lepton-number violating interactions. The basic
point is that non-perturbative baryon- and lepton-number violating (Sphaleron
induced) processes in the SM are thought to be in thermal equilibrium above the
electroweak breaking scale leading to the possible erasure of any pre-existing
baryon-number excess. These processes conserve $(B-L)$ and so, it is normally
argued, any baryon-number excess produced at an early epoch by a $(B-L)$
conserving interaction  will be erased. Thus $(B-L)$ violating processes at the
GUT scale are apparently needed if the baryon-number asymmetry is to survive.
Moreover, if new interactions are added which violate $(B-L)$ these, it seems,
in combination with the non-perturbative processes, will erase {\it all}
baryon-and lepton-number excess if they are in thermal equilibrium at some
temperature after baryogenesis.

In this paper we re-examine these constraints. We consider first the SM
sphaleron induced baryon- and lepton-number violating processes, including
corrections due to lepton-mass. Although small, these can lead to very
significant effects; for example an initial (flavour-dependent) lepton-number
asymmetry will {\it create} a baryon asymmetry through sphaleron processes
operating at low temperatures only when non-zero lepton masses are included.
Even a GUT-scale matter asymmetry created in a B=L channel {\it can } survive
the sphaleron interactions.  We also consider the constraints on the
introduction of baryon- and lepton-number interactions in supersymmetric
extensions of the SM. We extend the analysis of reference
\cite{campbell,fischler} of R-parity violating (\rpv) models to take account of
the lepton mass effects discussed above. As a result, we show the bounds on
baryon-number violating operators may be completely avoided. We also generalise
the analysis to allow for family dependence in the new \rpv couplings. As a
result we find bounds on the lepton-number violating couplings which are
substantially weaker than presented in \cite{campbell,fischler} and can be
completely avoided in a given flavour channel. These revised bounds allow the
possibility of finding R-parity violating signals in laboratory experiments. We
also consider the constraints that apply to supersymmetric GUTs when there may
be new sources of lepton number violation.

\section{Primordial Baryogenesis Constraints in the SM}
We begin by revising the analysis of the effect of sphaleron processes in the
SM
on a pre-existing baryon- and/or lepton-number asymmetry, taking account of
lepton mass effects on particle distributions in thermal equilibrium.

There is manifest evidence for baryon- and lepton-number violation ($\delb$ and
$\dell$) in cosmology: our solar system and very probably our galaxy contains
more matter than antimatter \cite{steigman,kolbturn}. For distant galaxies the
experimental evidence is inconclusive. If we assume that the observed galaxies
consist of matter and not antimatter then the presently observed
baryon-asymmetry of the universe is $n_B /s\approx  10{-10}$ \cite{kolbturn},
where $n_B$ is the net baryon-number density in the universe and $s$ is the
entropy density. The total lepton-number asymmetry is not observed directly and
can only be estimated \cite{kolbturn}. The apparent charge neutrality of the
universe \cite{lyttle} implies that $n_e/s\approx 0.85 \cdot (n_B/s)$
\cite{kolbturn}, where $n_e$ is the net electron-number density of the
universe.
No good estimate exists for the neutrino asymmetry; it is only bounded by the
limit on the overall energy density of the universe $n_{\nu_i}/s <4\cdot10{3}$
\cite{kolbturn}. There is a slightly stronger bound for the electron neutrino
asymmetry from nucleosynthesis $n_{\nu_e} /s <1\cdot10{3}$ \cite{olive}.

There are many suggestions for the origin of the baryon-asymmetry. These may
broadly be classified into early and late production depending on whether or
not
the baryon-asymmetry is produced far above the electroweak breaking scale (or
even input as an initial condition).  In the case of early baryogenesis there
are constraints on $\delb$ and $\dell$ operators, which may be added to the SM,
following from the requirement that they do {\it not} erase the baryon
asymmetry
in the subsequent evolution of the universe. We will discuss these constraints
in Section 4. In this section we concentrate on the constraints on the nature
of
the initial matter asymmetry which follow if it is not to be erased by the SM
sphaleron processes.

We shall do this by considering the number densities of the relevant particles
in terms of their chemical potentials, thereby taking into account all relevant
symmetries. In order to determine the full set of equations for the chemical
potentials, we must know which particle reactions are in thermal equilibrium.

\subsection{Equilibrium Conditions in the SM}
In the following we assume {\it early} baryogenesis and/or leptogenesis, except
where explicitly stated otherwise, and make no assumption that production
occurs
in a specific flavour channel. The electroweak interactions of the neutrinos
are
in thermal equilibrium down to about 1 MeV. Similarly the other electroweak
(gauge) as well as the strong interactions are in thermal equilibrium well
below
$T=\mw$. Next, we consider the possible $\delbi$ or $\delli$ interactions which
can be in thermal equilibrium below the GUT-scale, and explicitly list the
conserved quantum numbers.
\begin{itemize}
\item In the SM there are non-perturbative solutions of the electroweak
Lagrangian known as Sphalerons \cite{ht6,ht7,leptoruss}. The Sphaleron
interactions can be thought of as the creation out of the vacuum of an
\begin{equation}
\prod_i(u_Ld_Ld_L\nu_L)_i,\qquad i=1,\ldots,N
\label{eq:sphalinter}
\end{equation}
state \cite{harvey}. Here $u_L$, $d_L$ denote left-handed up- and down-like
quarks, $\nu_L$ denotes a left-handed neutrino, and $N=3$ is the number of
generations. These interactions violate baryon-number $B_i$ and lepton-number
$L_i$ but conserve the quantum-numbers
\begin{equation}
\third B-L_i,\quad B_i-B_j,
\label{eq:cons5}
\end{equation}
of which five are linearly independent. There are known Sphaleron solutions
which are in thermal equilibrium for temperatures between the W-boson mass and
the electroweak breaking scale $T_C\simeq 250\,GeV$.\footnote{The actual value
of the electroweak breaking scale depends on the Higgs mass; $T_{C}\approx
350GeV[1+(\frac{100GeV}{m_{\phi{0}}}){2}]{- 0.5}\cite{sw}$. We have here
assumed $m_{\phi}=100GeV$.} In addition, it is generally assumed that there are
further solutions which are in thermal equilibrium at temperatures $T>T_C$
\cite{tmw}. Thus we shall consider the interactions (\ref{eq:sphalinter}) to be
in thermal equilibrium for all temperatures $T\gsim\mw$ and the conserved
quantum numbers are given by Eq.(\ref{eq:cons5}).

\item The Yukawa interactions of the SM ($h_{i}\bar{l_{i}}l_{i}\phi{\star}$,
$h{\prime}_{ij}\bar{d}_{i}d_{j}\phi{\star}$,
$h_{ij}{\prime\prime}\bar{u}_{i}u_{j}\phi$),where $l$ and $\phi$ are the
charged leptons and the neutral Higgs scalar respectively,  are needed to give
mass to the leptons and quarks. The off-diagonal quark interactions violate all
three baryon quantum-numbers $B_i$, but they preserve the total baryon-number
$B$. Given the value of the Yukawa couplings we can determine whether the
interactions were in thermal equilibrium during some phase of the history of
the
universe. However, in the SM $h{\prime}$ and $h{\prime\prime}$ are not known,
since in electroweak processes we only observe the relative mixing between the
up-like and the down-like quarks and not the absolute mixing. However, the
off-diagonal elements of the Cabbibo-Kobayashi-Maskawa matrix provide a lower
bound for the absolute mixing and, conservatively, we will use this in our
estimates. The corresponding $\delbi$ interactions were in thermal equilibrium
if they violate the bound equivalent to Eq.(\ref{eq:barybound2}) ({\it cf.}
Section 4.2.2), corresponding to $h{\prime},h{\prime\prime}\gsim5 \cdot
10{-7} (M_{ Higgs}/1\,TeV){1/2}$. The smallest coupling is that between the
first and the third generation which, for $m_{top}\gsim150\,GeV$ is
\begin{equation}
h_{13}{\prime\prime}\gsim \frac{\sqrt{2}}{v}m_{top}\cdot V_{bu}
\approx  8\cdot 10{-4},  \label{eq:yukawa}
\end{equation}
The bound required for thermal equilibrium is satisfied for $h_{13}''$ and
therefore by all the other quark mass terms. We thus conclude that the
symmetries of Eq.(\ref{eq:cons5}) are reduced to the three conserved quantum
numbers
\begin{equation}
\third B-L_i.
\label{eq:cons3}
\end{equation}

\item The terms needed to give mass to leptons conserve all three lepton
quantum-numbers $L_i$, since the Higgs-lepton interactions are  flavour
diagonal
for massless neutrinos. Above $T_{C}$, even for massive neutrinos $\nu_e$, $
\nu_\mu$, and/or $\nu_\tau$, we do not expect the corresponding flavour
off-diagonal couplings to be in thermal equilibrium due to the smallness of the
neutrino masses and the associated Yukawa couplings. Below $T_{C}$ neutrino
oscillations could introduce significant lepton flavour violation. The
probability for oscillation is $sin{2}(2\theta)\, sin2 (\frac{\delta m{2}
t}{4T})$ where $\theta$ is the mixing angle between neutrino states, $\delta
m{2}$ is their mass${2}$ difference, t is the time between interactions and T
is the temperature which is much larger than the neutrino masses. Such flavour
violation will be in equilibrium only if $sin{2}(2\theta)\frac{\delta m{2}
M_{Planck}}{80T{3}} \ge 1$. At $T=M_{W}$ this corresponds to $sin(2\theta){2}
\delta m{2} \ge 4\cdot 10{6}eV{2}$. This bound must be checked for the case
of interest but we note that it is clearly not true for neutrino masses and
mixings needed to explain the solar neutrino defecit. We conclude that the
conserved quantum numbers of the SM and its modifications necessary to explain
the solar neutrino defecit are as given in Eq.(\ref{eq:cons3}).
\end{itemize}

\noindent Given the separate conserved lepton numbers it is clear from
Eq.(\ref{eq:cons3}) that a pre-existing baryon number survives provided there
is
a component produced in a $(\third B- L_{i})=0$ channel. However, as we show
below, even if no component is produced in a $(\third B-L_{i})=0$ channel a
GUT-scale matter genesis can survive in the SM.  In order to determine the  net
baryon-number, in particular whether it vanishes or not, we have found the
analysis via the chemical potentials to be the most convenient, thereby taking
into account all conserved quantum numbers. So we turn now to a discussion of
chemical potentials in the SM.

\subsection{Net Number Density}
The number density ${\cal N}_k(q)dq$ of particles of type $k$ with momentum
between $q$ and $q+dq$ is given by \cite{weinbook}
\begin{equation}
{\cal N}_k(q)dq= \frac{g_k} {2\pi2} q2 \frac{1} { \exp \left(
\frac {E_k(q)- \mu_k} {T} \right) \pm 1 }\,dq,
\end{equation}
where $E_k(q)\equiv (m_k2+q2){1/2}$ is the particle energy, $\mu_k$ is the
chemical potential, $g_k$ is the number of internal degrees of freedom and we
have set $\hbar=c=k=1$. The plus sign is for fermions and the minus sign is for
bosons. The chemical potential for particles and anti-particles is equal and
opposite so that we obtain for the net number density $n_k$ of a given particle
species
\begin{eqnarray}
n_k&=& \int_0\infty ({\cal N}_k(q)-{\bar {\cal N}}_k(q)) dq
\nonumber \\    &=& \frac{g_k} {2\pi2} \int_0\infty \left[
\frac{1}{ \exp \left(        \frac{E_k(q)-\mu_k}{T}  \right) \pm
1 } - \frac{1}{ \exp \left(        \frac{E_k(q)+\mu_k}{T} \right)
\pm 1 } \right] q2 dq \nonumber \\  & \approx & \frac{g_k}
{\pi2} T3 \left(\frac{\mu_k}{T}\right) \int_{m_k/T}\infty
    y \sqrt{y2- m_k2/T2} \frac{ey}{(1\pm ey)2} dy \nonumber
\\ &\equiv& \frac{g_k} {\pi2} T3 \left( \frac {\mu_k}{T}
\right) F_\pm( \frac          {m_k}{T})
\label{eq:chempot}
\end{eqnarray}
where in the penultimate line we have assumed $\mu/T\ll 1$ and made the
substitution $y=E_k(q)/T$. Thus we find that the number density of all
particles
is proportional to $\frac{g_k} {\pi2} T3 ({\mu_k}/{T})$ with a mass dependent
proportionality factor which we denote\footnote{For $m_k=0$, we find
$F_-(0)\approx 2.00003 F_+(0)$ in approximate agreement with \cite{harvey}.}
\begin{equation}
{F_\pm(\frac{m_k}{T})} = {F_\pm(0)} \alpha(\frac{m_k}{T}).
\end{equation}
For $T=\mw$ and $T_C$ we have listed the values of $1-\alpha$ in Table^1. In
the
following, we write equations for total charges ({\it e.g.} $Q$, $Q_3$, $B$,
and
$L$) in terms of the relevant $n_k$. Using Eq.(\ref{eq:chempot}) and dropping
the common factors $\frac{1}{\pi2} T2F_+(0)$ we will thus obtain equations in
terms of the corresponding chemical potential, $\alpha$, and $g_k$, where
$\alpha$ and $g_k$ are known. We shall solve the equations for the chemical
potentials;  the number density can always be obtained again via
Eq.(\ref{eq:chempot}). The mass correction factor was not included in previous
analyses and can be very important in some cases ({\it cf.} Sections 2.3 and
4.4).

\subsection{Chemical Potential Analysis}
In this section we revise the SM analysis of Ref.\cite{harvey} by taking mass
effects into account. We must first discuss the relevant chemical potentials in
the SM. In thermal equilibrium any charged particle can emit or absorb an
arbitrary number of photons; thus the chemical potential of the photon vanishes
at sufficiently high temperatures and densities. An analogous conclusion holds
for the $Z{0}$ boson (for $T\ge M_{Z}{0}$) and therefore the chemical
potentials of particles and antiparticles are equal and opposite. Similarly the
chemical potential of the gluon vanishes and thus different coloured quarks
have
equal chemical potentials. As discussed in Section 2.1, the off-diagonal Yukawa
interactions guarantee that the reaction chain $u\leftrightarrow
W{+}+s\leftrightarrow c\leftrightarrow t$ is in thermal equilibrium, so that
the chemical potential of all up-like quarks and all down-like quarks are
respectively equal. In the SM we are thus left with $3N+7$ chemical potentials
which we have summarised in Table^2 (N=3 is the number of families).
Furthermore
it is convenient to introduce the following notation
\begin{equation}
\begin{array}{rclrclrcl}
\alpha_i &\equiv&  \alpha({m_{e_i}}/ {T}), & \Delta_i &\equiv&
N-\sum_i \alpha_i, &&& \\ &&&&&&&& \\
\mu&\equiv& \sum_i\mu_i, &{\bar\mu} &=&\sum_i\alpha_i\mu_i, &
\Delta\mu &=& \mu-{\bar\mu}, \\ &&&&&&&& \\
\Delta_u &\equiv& N- \sum_i \alpha( {m_{u_i}}/ {\mw}), &
\Delta_d &\equiv & N- \sum_i \alpha( {m_{d_i}}/ {\mw}), &&&\\
&&&&&&&& \\ \alpha_- &\equiv&  \alpha( {m_{\phi-}} /{T}),
&\alpha_0 &\equiv& \alpha({m_{\phi0}} /{T}), &\alpha_W &\equiv&
\alpha({\mw}/{T}).
\end{array}
\label{eq:chemdef}
\end{equation}
In the massless limit $\Delta\mu=\Delta_i= \Delta_u= \Delta_d=0$, $\alpha_W=1$,
and $\alpha_- =\alpha_0=1$. At $T=\mw$ and for $m_{top}=150\,GeV$ we find
$\Delta_u \approx 0.38$, $\Delta_d \approx 5 \cdot 10 {-4}$, and $\Delta_i=
7\cdot 10{-5}$. In the following we shall therefore neglect $\Delta_d$ and
$\Delta_i$ compared to $\Delta_u$.

The electroweak interactions lead to the following $4+2N=10$ equilibrium
relations among the chemical potentials
\begin{equation}
\begin{array}{lll}
\mu_W &= \mu_-+ \mu_0 & (W- \lra \phi-+\phi0), \\
\mu_{dL} &= \mu_{uL}+ \mu_W  & (W- \lra {\bar u}_L + d_L), \\
\mu_{iL} &= \mu_i+ \mu_W   & (W- \lra {\bar\nu}_{iL}+ e_{iL}),\\
\mu_{uR} &= \mu_0+ \mu_{uL} & (\phi0 \lra {\bar u}_L +  u_R),
\\  \mu_{dR} &= -\mu_0+ \mu_W + \mu_{uL}&  (\phi0 \lra d_L +
{\bar d}_R), \\ \mu_{iR} &= -\mu_0+ \mu_W+\mu_i    & (\phi0 \lra
e_{iL}+ {\bar e}_{iR}). \end{array}
\label{eq:chemeq}
\end{equation}
Interactions in thermal equilibrium directly lead to (the above) equations for
the corresponding chemical potentials, independent of the mass corrections
$\alpha$ and internal degrees of freedom $g_k$ \cite{landau}.  However, the net
value of a quantum number, which we shall determine below, depends on the net
number density of the relevant particle species and thus depends on the product
of the chemical potential ($\mu$), the mass correction ($\alpha$), and the
number of internal degrees of freedom ($g_i$). Using the equations
(\ref{eq:chemeq}), we can express all of the chemical potentials in the SM in
terms of $(7+3N)- (4+2N)=3+N=6$ chemical potentials, which we chose to be
$\mu_W,\mu_0,\mu_{uL},$ and $\mu_i$. In the subsequent analysis $\mu_i$ only
appears in the combinations $\mu$ and $\Delta\mu$, we therefore will only have
5
independent chemical potentials.

We can now express the total electric charge $Q$ and the third component of
weak
isospin $Q_3$, as well as the total baryon number and lepton number in terms of
these 5 chemical potentials:
\begin{eqnarray}
Q&=& 2 (N-\Delta_u) (\mu_{uL} +\mu_{uR})- (N-\Delta_d) (\mu_{dL}+
\mu_{dR})-     \sum_i\alpha_i \cdot (\mu_{iL}+ \mu_{iR})
\nonumber \\  %
&&   -4\alpha_W \mu_W- 2b\alpha_-\mu_- \nonumber \\
&\approx&  2(N-2\Delta_u) \mu_{uL}  - 2(2N+ 2\alpha_W +b\alpha_-)
\mu_W   -2(\mu-\Delta\mu)  \nonumber \\
&&   +2(2N- \Delta_u + b\alpha_-)\mu_0,
\label{eq:charge} \\  && \nonumber \\
Q_3&=& \frac{3}{2} [(N-\Delta_u)\mu_{uL}- (N-\Delta_d)\mu_{dL}]+
\half\sum_i        (\mu_i-  \alpha_i \mu_{iL}) -4\alpha_W\mu_W
\nonumber \\ %
&&     -b(\alpha_0\mu_0 +\alpha_-\mu_-) \nonumber\\
&\approx&    -\frac{3}{2} \Delta_u \mu_{uL}- (2N+4\alpha_W
+b\alpha_-)\mu_W               +\half \Delta\mu + b(\alpha_- -
\alpha_0) \mu_0,  \label{eq:q3} \\ &&\nonumber \\
B&=& (N-\Delta_u) (\mu_{uL}+ \mu_{uR}) +(N-\Delta_d) (\mu_{dL}+
\mu_{dR}) \nonumber\\
&\approx&  (4N- 2\Delta_u )\mu_{uL} +2N \mu_W -\Delta_u
     \mu_0,\\ && \nonumber  \\
L&=& \sum_i [\mu_i+ \alpha_i(\mu_{iL} +\mu_{iR})] \nonumber \\
&=&  3\mu -2\Delta\mu +2N \mu_W -N\mu_0,
\end{eqnarray}
and $b$ is the number of Higgs doublets\footnote{In the SM b=1. In models such
as the supersymmstric version of the SM b=2 with different Higgs giving up and
down quarks their mass. However their chemical potentials are related so these
equations still apply(cf Section 4.1).} Within the SM we have for $T\gsim \mw$
one further equation for the chemical potentials due to the Sphaleron
interactions (\ref{eq:sphalinter})
\begin{eqnarray}
N(\mu_{uL}+2\mu_{dL})+\sum_i\mu_i&=&0, \\
\Rightarrow \quad N(3\mu_{uL}+2\mu_W)+\mu &=&0.
\label{eq:chemsphal}
\end{eqnarray}

At all temperatures $U(1)_Q$ is a good symmetry and therefore we must have $Q=0
$ \cite{lyttle}. Above the critical temperature of electroweak symmetry
breaking, $T_C$, we also have $Q_3=0$. For $T<T_C$ the latter no longer holds,
but then the non-vanishing vacuum expectation value of the neutral Higgs
implies
$\mu_0=0$. Thus for $T\gsim\mw $, we have 3 equations for the chemical
potentials, beyond those of Eq.(\ref{eq:chemeq}), for the five unknowns:
$\mu_{uL},\,\mu_W,\,\mu_0,\mu,$ and $ \Delta\mu$. Hence, we can write $B$ and
$L$ in terms of two chemical potentials, which we chose to be $\mu_{uL}$ and
$\Delta\mu$.

\subsubsection{$\bf T\gsim T_C$}
Above the scale for electroweak breaking the quarks and leptons, as well as the
W-boson are massless, $\Delta_u= \Delta_d= \Delta_i=0$, and
$\alpha_W=\alpha_i=1
$, the latter implying $\Delta\mu=0$. This is a further constraint on the
chemical potentials, giving 4 equations for 5 unknowns. Moreover,
$\alpha_-=\alpha_0$, but not necessarily unity.\footnote{Note that in this case
Eq.(\ref{eq:q3}) simplifies to $(2N+4+b\alpha_- ) \mu_W=0$, implying $\mu_W=0$
as it must be for $T>T_C$ \cite{harveyfoot}.} We can thus write $B$ and $L$ in
terms of $\mu_{uL}$ only
\begin{eqnarray}
B&=&4N\mu_{uL}, \label{eq:bsm} \\
L&=&- \frac {14N2+9Nb\alpha_-} {2N+b\alpha_-} \mu_{uL}.
\label{eq:baryhigh}
\end{eqnarray}
Thus, above $T_C$ the mass corrections do not modify the value of $B$ and only
very mildly modify $L$. From Eq.(\ref{eq:bsm}) and the observational value for
$n_B/s$ ({ \it cf.} Section 2.2.2) we obtain $\mu_{uL}\approx 10{-11}$ and
similar values for the other chemical potentials (except $\mu_W$). Combining
the
above two equations we obtain
\begin{equation}
B+L= -\frac {6N2+ 5Nb\alpha_-} {2N+b \alpha_-} \mu_{uL} = -\frac
{6N+5b\alpha_-} {22N+13b\alpha_-} (B-L),
\end{equation}
from which we conclude that a non-zero value for $B-L$ implies a non-zero value
for $B+L$, even though the $B+L$ violating Sphaleron interactions are in
thermal
equilibrium.

\subsubsection { $\bf\mw \lsim T \lsim T_C$}
In this temperature range all states have mass. Setting $\mu_0=0$ and using
Eq.(\ref{eq:charge}) we obtain
\begin{eqnarray}
B&=& \left(4N- 2\Delta_u +\frac {4N (2N- \Delta_u) } {2\alpha_W
+b\alpha_-} \right)     \mu_{uL} + \frac{2N} {2\alpha_W+ b
\alpha_-} \Delta\mu  , \\  %
L&=& -\left( 9N+ \frac {8N (2N- \Delta_u )} {2\alpha_W+b \alpha_-
}  \right)      \mu_{uL} - 2\left( 1+ \frac {2N} {2\alpha_W+
b\alpha_-}  \right) \Delta\mu, \\ %
B+L &=& -\left(5N+ 2\Delta_u+ \frac {4N(2N-\Delta_u)}
{2\alpha_W+b\alpha_-} \right) \mu_{uL} -\left( 2+ \frac {2N}
{2\alpha_W+ b\alpha_-}\right) \Delta\mu,\\ %
B-L &=&    \left(13N- 2\Delta_u+ \frac {12N(2N-\Delta_u)}
{2\alpha_W+ b\alpha_-} \right) \mu_{uL} +\left( 2+ \frac {6N}
{2\alpha_W+ b\alpha_-} \right)\Delta\mu ,  \label{eq:b-l}
\end{eqnarray}
In the massless limit Eqs.(\ref{eq:bsm}-\ref{eq:b-l}) agree with
\cite{nelson,harvey}. Again we find a non-zero value for $B+L$, which is not
washed out by the sphalerons. Inserting numbers we find the mass corrections
due
to the top quark modify previous results by only about $5\%$, the correction
due
to the Higgs and $W$-mass is about a factor of two. However, the more important
mass effect (due to $\Delta\mu\not=0$) is that $B+L$ is {\it not} proportional
to $B-L$ and therefore $B-L=0$ does not necessarily imply $B, B+L=0$.

This is contrary to previous results and is important, since it was  assumed
that due to the sphalerons one either requires early baryogenesis in a $B-L$
violating channel (this is for example {\it not} possible in the minimal
$SU(5)$
GUT) or  late baryogenesis at or below the electroweak phase transition.  In
the
following Section we show that this is not the case, {\it i.e.} even for
$B-L=0$, high-temperature matter-genesis {\it can} survive to the present.

\subsubsection {$\bf B-L=0$}
\label{sec-b-l}
In models where $B-L$ is conserved at all temperatures one must
impose the additional constraint $B-L=0$. For $T\gsim T_C$, this
additional equation for the chemical potentials immediately gives
$B=L= B+L =0$, as well as $\mu_{uL}=\mu_-= \mu=0$. However, we
do not necessarily have $\mu_i=0 $.

If the initial conditions are such that $L_i=0$, and if the only
lepton-number violating interactions are (lepton-flavor-blind)
sphaleron processes, then the $ \mu_i$ will be equal and the
vanishing of $\mu$ implies the vanishing of $\mu_i$ separately.
On the other hand, if there is a pre-existing asymmetry in a
given $L_i$ channel, then $L_i- L_j$ conservation, as implied by
Eq.(\ref{eq:cons3}), means that the vanishing of $L$ derived here
can come about only through a cancellation between different,
non-zero, $\mu_i$. In that case we find, when considering the
low-temperature region, that the non-zero $\mu_i$ imply a  non-zero
$\Delta\mu$ which will {\it regenerate} the lepton- and
baryon-asymmetry, similar to models discussed in
\cite{leptoruss,fukugita}. This follows because, imposing $B-
L=0$, we can express $\mu_{uL}$ by $\Delta\mu$ using
Eq.(\ref{eq:b-l}) and then obtain
\begin{equation}
B = \left[ - \frac { \left(4N- 2\Delta_u +\frac {4N (2N-
\Delta_u)  } {2\alpha_W +b\alpha_-} \right) \left( {2{\alpha_W+b
\alpha_-}+ {3N}}\right)} {(13N-2 \Delta_u) ({\alpha_W+
\frac{b}{2}\alpha_-}) + {6N(2N-\Delta_u)}} + \frac{2N}
{2\alpha_W+ b\alpha_-} \right] \Delta\mu,
\label{eq:barylow}
\end{equation}
as well as $B=L=\half(B+L)$. In this case $\Delta\mu \not= 0$,
since $\mu_i \not=0$ and the baryon number  reappears {\it due}
to the sphaleron interactions converting $L_i$ number excess into
$B$ excess. (Note that in the case $\mu_i=0$, discussed above,
$\Delta\mu=0$ and no baryon number is recreated when $B-L=0$.)
Thus we find the surprising result, that even when $B-L=0$ for
all temperatures we can have non-vanishing $B,L$ and $B+L$.

We note that the result of Eq.(\ref{eq:barylow}) also implies
that any low-temperature baryogenesis ($T<T_C$) is not washed out
by the sphalerons, even if $B-L=0$, as long as it is not family-symmetric in
the $B+L_i$ channels. Thus if it is possible to
create baryogenesis at the electroweak phase transision, and if
it is family {\it asymmetric} in the lepton channels (for example
due to lepton mass effects) then, contrary to common wisdom, the
sphalerons will not subsequently erase it, even if they are in
thermal equilibrium below the reheating temperature.

We now extend our analysis to include supersymmetry.

\section{The MSSM and its \rpv\ Extensions}
\subsection{Introduction}
The supersymmetric extension of the SM introduces new scalar
matter states, the squarks and sleptons, which carry the same
internal quantum numbers as their fermion partners, the quarks
and leptons. These new states also give rise to new possibilities
for baryon- and lepton-number violating processes. In order to
specify the supersymmetric model it is necessary to detail the
Yukawa and scalar interactions involving these new states
together with the interactions involving the Higgs bosons and the
Higgsinos, their supersymmetric partners. These are specified by
the superpotential, W, and to obtain the usual couplings of the
SM giving rise to quark and lepton masses we need the terms
\begin{equation}
W=h_i[L_iH_2{\bar E}_i]_F+ h'_{ij}[Q_iH_2{\bar D}_j]_F+
h''_{ij}[Q_iH_1{\bar U} _j]_F,
\label{eq:higgs}
\end{equation}
where $L$ and ${\bar E}$ ($Q$ and ${\bar U},{\bar D}$) are the
(left-handed) lepton doublet and the anti-lepton singlet (quark
doublet and anti-quark singlet) chiral superfields respectively
and $H_{1,2}$ are the Higgs chiral  superfield doublets.
$i,j=1,2,3$ are generation indices and $h,h'$, and $h''$ are
dimensionless Yukawa coupling constants. Note that the lepton-Higgs
interaction is generation diagonal since in the SM the
neutrino masses vanish. The quark-Higgs couplings include non-zero
off-diagonal terms due to the Kobayashi-Maskawa mixing of
the quarks ({\it cf.} Section 2.1).

These couplings are the only ones introduced in defining the MSSM
which has been extensively analyzed as the paradigm for a
supersymmetric theory \cite{susy}. The theory thus defined
conserves R-parity defined by  \begin{equation}
R_p=(-){2S+3B+L},
\end{equation}
where $S,B,$ and $L$ are the spin, the total baryon, and the
total lepton quantum-numbers, respectively.

The terms in Eq.(\ref{eq:higgs}) are of course allowed by the
$SU(3)\otimes SU( 2)\otimes U(1)$ gauge invariance but they are
not the only such terms. The most general trilinear terms consist
of those above together with the terms
(operators) \cite{gaugterms}
\begin{equation}
W_{\not\!\!R_p}=\lam_{ijk} [L_{i}L_{j}\bar{E}_{k}]_F+ \lam'_{ijk}
[L_{i}Q_{j}  \bar{D}_{k} ]_F + \lam''_{ijk} [\bar{U}_{i}
\bar{D}_{j}\bar{D}_{k}]_F,  \label{eq:rpviolat}
\end{equation}
where $i,j,k$ are generation indices and $\lam,\lam'$, and
$\lam''$ are dimensionless Yukawa coupling constants. (In
addition there may be bilinear terms
[$L_{i}H_{1}]_{F}$ which, however, can be eliminated by
redefining the lepton and Higgs fields provided we include terms
of the form of the first two in Eq.(\ref{eq:rpviolat}); so we do
not need to consider them separately.)

The terms of Eq.(\ref{eq:rpviolat}) violate baryon- and lepton-number
and they generate an unacceptably large amplitude for
proton decay suppressed only by the inverse squark mass squared.
For this reason a discrete symmetry called matter parity was
introduced which forbids such new terms.
Under this symmetry the quark and lepton superfields appearing
in the superpotential change sign while the Higgs superfields are
left invariant. Thus the terms of Eq.(\ref{eq:rpviolat}) change
sign under this symmetry and are forbidden while the terms of
Eq.(\ref{eq:higgs}) are invariant and allowed. Using this
symmetry the couplings of the MSSM are the only ones allowed and
the new supersymmetric states only couple in pairs in accord with
the R-parity of the model.

However, the origin of the discrete matter parity is not clear
at the level of the MSSM; presumably it comes from the underlying
unified theory. Before the advent of string theory it was thought
likely that quarks and leptons would transform in the same way
in a Grand Unified extension of the SM and this suggested that
imposition of matter parity was the most reasonable way to
supersymmeterise the SM. However, string theories do not in
general have discrete symmetries which treat quarks and leptons
in the same way. It is easy to write down alternative symmetries
to matter parity which allow for a different set of couplings.
For example, if only the quark superfields change sign then only
the last (baryon-number violating) term  in
Eq.(\ref{eq:rpviolat}) is forbidden. This is consistent with
nuclear decay experiments, provided at least one of the last two
vertices of Eq.(\ref{eq:rpviolat}) is absent the nucleon may be
sufficiently stable to be consistent with present
limits.\footnote{If the first and the third operators are
present, proton decay will proceed but via a multiparticle
amplitude which is considerably suppressed. In practice, however,
it is safest to suppress either the first or {\it both} the
second and third operators to eliminate proton decay at an
unacceptably fast rate.} It is encouraging that in specific
examples of compactified string theories just such discrete
symmetries emerge \cite{rpstring}, they show that the MSSM is not
the only viable way to construct a minimal supersymmetric
extension of the SM. This  observation suggests that equally
reasonable are alternative minimal
supersymmetric models in which any subset of the terms of
Eq.(\ref{eq:rpviolat}) which does not give rise to nucleon decay
is included.\footnote{We note that it is also possible to
construct supersymmetric, \rpv\ GUTs \cite{RP1,flipped}.} What
is clearly needed is a criterion to decide between the two
possibilities.

\medskip
Recently it has been shown \cite{ibanross}, that the allowed
forms of discrete symmetries, which are needed to inhibit nucleon
decay in supersymmetric versions of the SM, are strongly
constrained by the requirement of anomaly cancellation if they
come from underlying gauge symmetries, broken at some high
scale.\footnote{It has been argued that only such ``discrete
gauge symmetries'' can survive large gravitational corrections
and thus be useful in inhibiting nucleon decay.} A survey of all
possible discrete symmetries of low dimension ($Z_{N},N\le 4$)
showed that there are only two viable ``discrete gauge
symmetries'' with the minimal light matter content of the SM;
namely \rp\ and a new $Z_{3}$, called ``baryon-parity'', which
allows the class of lepton-number violating \rpv\ operators ({\it
cf.} Eq.(\ref{eq:rplep})). Indeed, the latter has the advantage
of eliminating the dimension-5 operators which generate nucleon
decay and it was argued that it deserves to be considered at
least as seriously as \rp\ when constructing a supersymmetric
version of the SM.

\subsection{Constraints}
Given this ambiguity in constructing the supersymmetric version
of the SM it is  clearly of interest to look for constraints on
such extensions. In the following we shall consider the general
set of \rpv\ models which do not generate unacceptable levels of
proton decay. Thus we consider the $\delb$ operators
\begin{equation}
\lam''_{i j k} [\bar{U}_{i} \bar{D}_{j}\bar{D}_{k}]_{F},
\label{eq:rpbary}
\end{equation}
separately from the $\dell$ operators
\begin{equation}
\lam_{i j k} [L_{i} L_{j} \bar{E}_{k}]_F + \lam'_{i j k} [L_{i}
Q_{j} \bar{D}_{k}]_F,
\label{eq:rplep}
\end{equation}
and we shall assume that they are not simultaneously present in
the Lagrangian. Let us first comment on the likely family
structure of new terms of Eq.(\ref{eq:rpviolat}). Of the allowed
couplings chosen consistent with nucleon stability, the most
likely case is that one of them dominates because the new
couplings are of the same character as the normal Yukawa
couplings responsible for quark and lepton masses and these span
a vast range in magnitude. It is also likely that all operators
related to the dominant one by mixing of the quark and lepton
(super) states will be present since, even if a symmetry is
present at the current eigenstate level which allows only one
family structure of operator, this will be broken by the mass
matrix elements mixing different families. However, this mixing
is known to be quite small in the quark sector and so the
coefficients of the related operators may naturally be quite
small. For this reason we think it necessary to analyze the
bounds allowing for substantial differences in the couplings in
Eq.(\ref{eq:rpviolat}). This leads to important modifications of
the bounds following from the requirement of non-erasure of
primordial baryogenesis as we will discuss in Section 4.4.

\subsubsection{Experimental Constraints}
Present experimental upper bounds following from the virtual
effect of the new supersymmetric states in processes involving
SM states limit many of the \rpv-Yukawa coupling constants to be
less than or of the order of $0.2$ -$0.01\,\left({m_{\tilde
f}}/{100\,GeV} \right)$ \cite{vernon,rpcoll}, where $m_{\tilde
f}$ is the scalar fermion mass relevant to the process being
considered. However, the bounds are family dependent with weaker
bounds for the operators involving the heavier families; indeed
some couplings are not bounded  at all.  A limited set of \rpv-operators
can be directly tested through  s-channel single
supersymmetric particle production, {\it e.g.} the operators
$[L_1Q_i{\bar D}_j]_F$ through squark production at HERA
\cite{butter} and slepton production at hadron colliders
\cite{rp3}. In these cases the coupling constants can be probed
down to $\lam \approx5\cdot10{-3}$, for sfermion masses in the
kinematic range of the collider. Therefore we expect that some
of the bounds should be  considerably improved in the near future
or indeed supersymmetry with \rpv\  will be found.

\subsubsection{Cosmological constraints}
Recently, there has been much work in trying to obtain bounds on
\rpv-couplings from cosmology
\cite{campbell,fischler,boqsal,barbmass} since over the large
space-time scales of the universe the small \rpv-couplings can
lead to visible effects. However, many of the bounds involve
products of \rpv-couplings and given our expectation that the
various couplings are likely to differ considerably in magnitude
these do not rule out the possibility that there will be some
large couplings accessible to future collider tests. For this
reason we concentrate here on constraints following from the need
to produce a baryon-number excess which do limit individual
couplings and therefore are of some importance when considering
the possibilities for experimental tests of \rpv\  models.

Now, one might have thought that if a $\delb$ operator comes into
thermal equilibrium after baryogenesis it will erase the baryon-asymmetry,
no matter what other interactions are present. One
could then interpret the  out-of-equilibrium condition for
interactions based on the $\delb$,  \rpv-operators of
Eq.(\ref{eq:rpbary}) as a strict bound based on early
baryogenesis. In Ref.\cite{boqsal} this was analysed for the
$\delb$, $2\ra2$ interaction shown in Fig.^1a obtaining the bound

\begin{equation}
\lam''\lsim 10{-6}\cdot \left( \frac {{\tilde m}} {100\,GeV}
\right){1/2}. \label{eq:limsal}
\end{equation}
No account was taken of non-perturbative electroweak effects
which may violate  B and L at high temperatures. In recent work
\cite{campbell,fischler} they have been included and the
analogous bounds to Eq.(\ref{eq:limsal}) for {\it all} \rpv-couplings,
including the $\dell$ operators (\ref{eq:rplep}), were
obtained \begin{equation}
\lam,\lam',\lam'' \lsim 10{-8}.
\label{eq:limits}
\end{equation}
These were based on a dimensional analysis of the high-temperature
limit of the $\delb$ and $\dell$, $2 \ra2$ processes
shown in Figs.^1a,b,c (therefore the discrepancy with
Eq.(\ref{eq:limsal}) above) and were also interpreted as bounds
from early baryogenesis. As we show below, including the $2\ra1$
processes shown in Figs.^2a,b,c and performing a complete
calculation of the cross sections gives the slightly stronger
(compared to Eq.(\ref{eq:limsal})) out-of-equilibrium condition
\begin{equation}
\lam, \lam', \lam'' \lsim 5\cdot 10{-7} \left( \frac{{\tilde m}}
{1\,TeV}  \right) { 1/2}.
\label{eq:limits2}
\end{equation}
If the conditions of
Eqs.(\ref{eq:limsal},\ref{eq:limits},\ref{eq:limits2}) were
indeed strict bounds, then \rpv\ is of no relevance for collider
physics. Using the above limiting value for the coupling, the
lifetime of the lightest supersymmetric particle (assuming it is
a photino) is given by \cite{Dawson}  \begin{equation}
\tau_{{\tilde\gamma}} \approx 2\cdot 10{-6} (M_{{\tilde f}}/
200^GeV)4  (50^GeV/M_{\tilde\gamma})5\,sec,
\label{eq:lifetime}
\end{equation}
where $M_{\tilde f}$ is the mass of the relevant scalar fermion.
Therefore only for a large photino mass and small sfermion masses
will the photino decay within the detector
($\tau_{{\tilde\gamma}}\lsim 10{-9}\,sec$). If it decays outside
the detector most signals are identical to the case of conserved
R-parity. The remaining distinctive \rpv\  possibility of single
production of supersymmetric particles requires coupling
strengths of order $10{-3}$ or larger in order to  lead to
visible effects \cite{butter}, and thus would also be irrelevant.
In  contrast, in the evolution of the universe a photino lifetime
of $10{-6}\,sec$ or more can still lead to observational effects
\cite{boqsal}.

However, as we discuss below ({\it cf} Sections 4.3, 4.4), the
situation is more complicated and the bounds of
Eqs.(\ref{eq:limsal},\ref{eq:limits},\ref{eq:limits2}) are {\it
not} strict bounds. A generalisation of the discussion presented
above for the SM shows that even with (non-SM) $\delb$
interactions in thermal equilibrium there are further remnant
symmetries which protect the baryon-asymmetry. The requirement
that the $\delb$ and $\dell$ interactions are out of thermal
equilibrium for all temperatures below the scale of baryogenesis,
$T<T_{baryog}$ ($\mw<T<T_{baryog}$ in the case where Sphalerons
are included), is merely a {\it sufficient} but {\it not a
necessary} condition for an early baryon-asymmetry to survive
until today.

\section{Baryogenesis Constraints in the MSSM and its \rpv\
extensions}  Given the severity of the bounds
(\ref{eq:limsal},\ref{eq:limits},\ref{eq:limits2}) following from
baryogenesis it is important to discuss their range of
applicability in particular to determine whether laboratory
searches for \rpv\ couplings are futile. The first important
assumption leading to  Eq.(\ref{eq:limits}) is early
baryogenesis; an obvious caveat is therefore that baryogenesis
did not occur until a late phase in the evolution of the
universe. There are many proposed mechanisms for late
baryogenesis \cite{dolgov} one of which even involves the \rpv-operators
themselves \cite{dimhall}. These would render the above
bounds obsolete.

Even if we restrict ourselves to models with early baryogenesis
there are important cases  in which the bounds of
Eq.(\ref{eq:limits}) do not apply. We shall rederive the bounds
from baryogensis on all the operators
(\ref{eq:rpbary},\ref{eq:rplep}) modifying previous analyses
\cite{boqsal,campbell,fischler} in three important  aspects.

The first modification amounts to including an analysis of the
chemical  potentials; thereby taking into account all the
relevant symmetries.

The second and third modifications correspond to relaxing two
important assumptions which were made in \cite{campbell} and
which need not be the case in realistic models. These assumptions
were
\begin{description}
\item[(A1)] The $\dell$ couplings in Eq.(\ref{eq:rplep}) are
family symmetric, so all three lepton quantum-numbers are
violated to the same degree, {\it e.g.} the ratio of $\lam_{122}$
which violates $L_1$ to $\lam_{211}$ which violates $L_2$ is of
order one: $\frac{\lam_{122}} {\lam_{211}} \sim \oc(1)$.
\item[(A2)] The present baryon-asymmetry is due to {\it
baryogenesis} at the GUT-scale, as opposed to leptogenesis, for
example.
\end{description}
We will see that (I) taking account of the relevant symmetries
via the chemical potential analysis eliminates the bounds on the
$\delb$ operators completely in the case where there are no
Sphaleron interactions present (this is just the case of Section
2.3) as well as in the case of late baryogenesis, but retains the
bound (\ref{eq:limits2}) in the case of early baryogenesis and
Sphaleron interactions in thermal equilibrium. (II) Abandoning
(A1) eliminates the bounds on the $\dell$ couplings in a given
flavor channel; (III) Abandoning (A2) eliminates the bounds on
the $\delb$ operators completely, while leaving the results for
the $\dell$ operators untouched.

 We start with the interactions of the MSSM and then study the
\rpv-interactions. In Section 4.3 we will then solve the
equations for the chemical potentials in order to determine the
resulting low-temperature total baryon- and lepton-number.
\begin{itemize}
\item  The supersymmetric gauge (and gaugino) interactions will
be in thermal equilibrium until the temperature drops below the
relevant SUSY mass scale, as discussed in Section 2.1 for the SM.
Similarly, the supersymmetric quark-Higgs  Yukawa interactions
(off-diagonal and diagonal) of Eq.(\ref{eq:higgs}) are in thermal
equilibrium and the symmetries of the MSSM are those of Eq(4).
\end{itemize}
\noindent In coming to this conclusion we assume there are no
additional sources of slepton mass other than that in
Eq.(\ref{eq:higgs}) and a flavour blind mass coming from SUSY
breaking in the hidden sector. This is expected in most
superstring models which do not have a stage of Grand Unification
but is not in general the case in SUSY GUTs where new
interactions at the GUT scale appear. We will discuss this in
Section^\ref{sec-susy}.

\subsection{Chemical Potentials in the MSSM}
In the MSSM there are two doublets of Higgs fields. However they
are strongly coupled and as a result their chemical potentials
are related (their sum vanishes). This means the relations
involving Higgs chemical potentials reduces to those found in the
standard model discussed above where the $\phi$ and $\phi{*}$
had equal and opposite chemical potentials.

The supersymmetric partners of the gluons, the $W0$, and the
$B0$, respectively, are Majorana fermions. One might therefore
expect their chemical potential to vanish. However, as shown in
\cite{langacker}, this is not necessarily so. The  distinct
helicity states ($h=\pm1$) of Majorana fermions can be thought
of as particle and anti-particle, and provided the helicity-flip
interactions are out of thermal equilibrium, an effective
chemical potential can be introduced. The helicity-flip rate,
resulting from the Majorana mass terms, is $O(m_i2/E_i2)$ and
thus, indeed, we expect it to be in thermal equilibrium in the
case of the gluinos and neutralinos (in contrast to the case of
light Majorana neutrinos for example). Therefore the chemical
potentials of the gluinos, and the neutralinos vanishes.
Furthermore, provided for example the interactions ${\tilde
f}\lra f+(gluino, neutralino)$ are in thermal equilibrium, {\it
i.e.} for $T>m_{\tilde f}$, we conclude that the chemical
potentials of the fermions and their corresponding scalar fermion
superpartners are equal, $\mu_{ \tilde f}= \mu_f$. The analogous
conclusions hold for the charginos: $\mu_{ \tilde W} =\mu_W $ and
$\mu_{{ \tilde \phi}-}= \mu_-$. Thus, above the supersymmetric
thresholds, the extension of the SM analysis to the MSSM analysis
does not lead to the introduction of further chemical potentials
and the SM analysis of Section 2.3 remains valid,\footnote{At
very high temperatures compared to the SUSY masses and/or the
SUSY-breaking scale the helicity flip rate is suppressed and
these conclusions do not hold \cite{luis}. One must then indeed
consider non-zero chemical potentials for the gluinos and
neutralinos. However, this does not affect the conclusions of our
work.}\, however with at least 2 Higgs doublets $b\geq2$. Below
the supersymmetric thresholds we expect the sparticles (except
the LSP) to annihilate and drop out of equilibrium and therefore
also not contribute to the analysis. In the MSSM we are thus left
with the $7+3N=16$ chemical potentials of the SM, shown in
Table^2, which reduce to the 5 independent chemical potentials
$\mu_{u_{i}},\mu_{W},\mu_{0},\mu,\Delta \mu$ after using
Eqs(\ref{eq:chemeq}).

Given these chemical potentials we may express the supersymmetric
expressions for the total electric charge $Q$ and the third
component of weak isospin $Q_3$, as well as the total baryon
number and lepton number, applicable above the supersymmetric
threshold when the supersymmetric states are in equilibrium.
\begin{eqnarray}
Q&=& Q_0+Q_S\nonumber \\
Q_0&=& 2 (N-\Delta_u) (\mu_{uL} +\mu_{uR})- (N-\Delta_d)
(\mu_{dL}+ \mu_{dR})-     \sum_i\alpha_i \cdot (\mu_{iL}+
\mu_{iR}) \nonumber \\  %
&&   -4\alpha_W \mu_W- 2b\alpha_-\mu_- \nonumber \\
Q_S&=& 4 (N-\Delta_{\tilde{u}}) (\mu_{uL} +\mu_{uR})-2 (N-
\Delta_{\tilde{d}}) (\mu_{dL}+ \mu_{dR})-   2
\sum_i\tilde{\alpha}_i \cdot (\mu_{iL}+ \mu_{iR}) \nonumber \\
&&   -\alpha_{\tilde{W}} \mu_W- b\tilde{\alpha}_-\mu_- \\  %
&& \nonumber \\
Q_3&=& (Q_3)_0+ (Q_3)_S\nonumber \\
(Q_3)_S &=& \frac{3}{2} [(N-\Delta_u)\mu_{uL}- (N-
\Delta_d)\mu_{dL}]+ \half\sum_i        (\mu_i-  \alpha_i
\mu_{iL}) -4\alpha_W\mu_W \nonumber \\ %
&&     -b(\alpha_0\mu_0 +\alpha_-\mu_-) \nonumber\\
(Q_3)_S&=& 3 [(N-\Delta_{\tilde{u}})\mu_{uL}- (N-
\Delta_{\tilde{d}})\mu_{dL}]+ \sum_i
(\alpha_{\tilde{\nu}_{i}}\mu_i-  \tilde{\alpha}_i \mu_{iL}) -
\alpha_{\tilde{W}}\mu_W \nonumber \\ %
&&     -\frac{1}{2}b(\tilde{\alpha}_0\mu_0 +\tilde{\alpha}_-\mu_-
)\\
&& \nonumber \\
B&=& B_0+B_S\nonumber \\
B_0&=& (N-\Delta_u) (\mu_{uL}+ \mu_{uR}) +(N-\Delta_d) (\mu_{dL}+
\mu_{dR}) \nonumber\\
B_S&=& 2 (N-\Delta_{\tilde{u}}) (\mu_{uL}+ \mu_{uR}) +2(N-
\Delta_{\tilde{d}}) (\mu_{dL}+ \mu_{dR}) \\
&&\nonumber \\
L&=& L_0+L_S\nonumber \\
L_0&=& \sum_i [\mu_i+ \alpha_i(\mu_{iL} +\mu_{iR})] \nonumber \\
L_S&=& 2\sum_i [\alpha_{\tilde{\nu}_{i}}\mu_i+
\tilde{\alpha}_i(\mu_{iL} +\mu_{iR})]
\label{eq:susychem}
\end{eqnarray}
where tildes denote the supersymmetric analogues of the non-supersymmetric
quantities defined above and we have had to introduce an $\alpha$ for the
sneutrinos which are massive. The
two components of each quantum number correspond to the non-supersymmetric
contribution (labelled 0) and the supersymmetric
contribution (labelled S).

\subsection{Equilibrium Conditions for \rpv-Operators}
In order to determine in general when early baryogenesis is
possible in \rpv-{\it{models}} we will need to know when the
\rpv-interactions are in thermal equilibrium, and thus which
chemical potential equations are relevant.  In order to determine
the out-of-equilibrium conditions for the \rpv-interactions we
consider both the $2\ra2$ interactions of Fig.^1 and the $2\ra1$
interactions of Fig.^2. We assume that the relevant scalar
fermion mass $m_{\tilde f}>\mw$.

\subsubsection{$2\ra2$ Processes}
For the $2\ra2$ process of Fig.^1a the annihilation rate is given
by \cite{weinberg}
\begin{equation}
\Gamma\simeq \frac{2\alpha \lam''2e_q2}{\pi3}
\frac{T5}{\left[ T2+\msq2\right]2},
\label{eq:rate2}
\end{equation}
where $e_q$ is the charge of the squark in units of the electron
charge, $\alpha $ is the fine-structure constant, and we have
neglected final state masses. ${ \tilde m}$ is the mass of the
s-channel squark. The Hubble expansion rate is given by
\begin{equation}
H=1.66\,\frac{T2 D{1/2}}{\mpl}\simeq \frac{20 T2}{\mpl},
\label{eq:hubble}
\end{equation}
where $\mpl=1.2\cdot 10{19}\,GeV$ is the Planck mass and $D$ is
the number of  effective degrees of freedom. The annihilation
rate is out of thermal  equilibrium  for $\Gamma/H<1$. The ratio
of the annihilation rate  (\ref{eq:rate2}) to the expansion rate
is given by
\begin{equation}
\frac{\Gamma}{H}\simeq\frac{\alpha\lam''2e_q2}{10\pi3}\,\frac
{T3\mpl}{[T2+{ \tilde m}2]2},
\label{eq:ratio1}
\end{equation}
and $\Gamma/H$ is maximal at $T_{max}= \sqrt{ 3}\,{\tilde m}$.
The out-of-equilibrium condition is $\Gamma/H(T_{max})<1$ and
translates to   \begin{equation}
\lam''\lsim 5\cdot10{-6}\left(\frac{\msq}{1\,TeV}\right){1/2},
\label{eq:barybound1}
\end{equation}
which agrees with Eq.(\ref{eq:limsal}) of Ref.^\cite{boqsal}.
However, as we now show, the $2\ra1$ processes of Fig.^2 lead to
even stricter conditions.

\subsubsection{$2\ra1$ Processes}
The $2\ra1$ processes are only relevant for $T>{\tilde m}$; for
lower temperatures the initial state quarks do not have
sufficient energy and the processes are exponentially suppressed.
The phase space integral of the  amplitude squared for the
$2\ra1$ annihilation process is given by  \begin{equation}  \int
dPS |{\cal M}|2 =c_f\,\frac{\pi\lam''2} {4}\,
\msq2\,\delta(s-\msq),  \label{eq:sigma1}
\end{equation}
where $c_f=\frac{2}{3}$ is the colour factor, $\lam''$ is the
relevant  Yukawa coupling and $s$ is the center-of-mass energy
squared. After thermally averaging over the momenta of the
incoming quarks the annihilation rate is given by
\begin{equation}
\Gamma=<\sigma\cdot v_{rel}>\cdot
n_0=\frac{\lam''2}{9\pi\zeta(3)}\,
\frac{\msq2}{T}\,f(\frac{T2}{\msq2}),
\label{eq:rate1}
\end{equation}
where we have used the equilibrium number density $n_0$ as an
approximation and $\zeta(3)=1.202$ is the Riemann Zeta-function.
The function f is given by  \begin{equation}
f(x)=\int_0\infty  \frac{\ln(1+e{-x/4y})}{ey+1}\,dy.
\label{int}
\end{equation}
It is well approximated by: $f(x)=a+b\ln(x2)$, where $a=0.27$
and $b=4.3\cdot 10{-3}$, and thus only has a very weak
temperature dependence. The ratio of the annihilation rate
(\ref{eq:rate1}) to the expansion rate (\ref{eq:hubble}) is given
by
\begin{equation}
\frac {\Gamma} {H} \simeq \frac {\lam''2} {180 \pi\zeta(3)}\,
\frac {\msq2\mpl} {T3} \, f(\frac{T2}{\msq2}).
\label{eq:ratio2}
\end{equation}
The annihilation rate is out of thermal equilibrium for
$\Gamma/H<1$. $\Gamma/H$ is at a maximum as a function of $T$ at
$T= \msq$ and we obtain the condition  \begin{equation}
\lam''\lsim \left( \frac {180\pi\zeta(3)} {f(1)} \frac {\msq}
{\mpl} \right) {1/2} \lsim 5\cdot 10{-7} \left( \frac {\msq}
{1\,TeV} \right) {1/2},  \label{eq:barybound2}
\end{equation}
which is stronger than Eq.(\ref{eq:barybound1}) by an order of
magnitude.\footnote{We note that there is some ambiguity in the
choice of  temperature, when the $2\ra1$ process goes out of
equilibrium. We have chosen the scale $T=\msq$, which leads to
the weakest bound. If instead we choose $T= \msq/2$, for example,
the bound (\ref{eq:barybound2}) is strengthened to
$\lambda''\lsim 1.8\cdot 10{-7} (\msq/1\,TeV){1/2}$.} If we
wish to only  demand out-of-equilibrium above a certain
temperature $T>\msq$, for example the critical temperature of
electroweak symmetry breaking, we obtain the slightly weaker
bound
\begin{equation}
\lam''\lsim \left( \frac {180\pi\zeta(3)} {f(T2_C/{\tilde m}2)}
\frac {T3_C}  {\mpl\msq2} \right) {1/2} \lsim 6\cdot 10{-7}
\left(\frac {100\,GeV} {\msq} \right).
\label{eq:barybound3}
\end{equation}

\subsection{Baryogenesis constraints on $\delb$, \rpv\ Operators}
Here we discuss the case of the $\delb$ couplings
(\ref{eq:rpbary}), and assume vanishing $\delli$ couplings
(\ref{eq:rplep}), {\it i.e.} $\lam,\lam'\equiv0$.

\subsubsection{No Sphalerons}
First, let us consider the simplest case with {\it no} Sphaleron
interactions in thermal equilibrium for this clearly exhibits the
pitfalls that may be met unless the analysis employing chemical
potentials is used. (It is also the relevant scenario for
$T<\mw$.) We obtain an additional chemical potential equation
through the operators (\ref{eq:rpbary}), if the new $\delb$ terms
are in thermal equilibrium
\begin{eqnarray}
\mu_{uR}+ 2\mu_{dR}&=&0, \\
\Rightarrow\quad 3\mu_{uL}+ 2\mu_W -\mu_0 &=&0.
\label{eq:chembary}
\end{eqnarray}
Dropping the sphaleron equation (\ref{eq:chemsphal}) and
including the previous equation we have 3 constraint equations.
For temperatures far above the supersymmetric masses and the
electroweak breaking scale the mass corrections may be ignored
and we find (for N=3)\newline

\vspace{0.5cm}

$\bf T\gsim T_C,\,\, T>> M_{SUSY}$
\begin{eqnarray}
B&=& 36 \mu_{uL} \label{eq:nosphal1} \\
L&=& 126 \mu_{uL},
\label{eq:nosphal2}
\end{eqnarray}
In the case of complete absence of Sphaleron interactions for all
temperatures we obtain the surprising results
(\ref{eq:nosphal1}-\ref{eq:nosphal2}) that,  even though a baryon number
violating
interaction is in thermal equilibrium, there are non-trivial
solutions for $B$ and $L$ which relate $B$ to $L$. Thus a non-zero Baryon
number excess may persist in the presence of the
$\Delta B\ne 0$ operators, a result in
disagreement with Eq(\ref{eq:limsal}) from Ref.\cite{boqsal};
{\it i.e.} there is no bound in general. The reason for this is
that charge neutrality relates baryon and lepton abundances; an
initial lepton asymmetry is not erased (in the absence of
sphalerons) by a B violating interaction and in thermal
equilibrium the resulting excess of charged leptons of a given
charge must be balanced by an excess of charged baryons of the
opposite charge. The analysis in terms of chemical potentials
nicely accounts for this.

The non-trivial solution for B and L persists when mass effects
are included and can be calculated from Eq(32)-(36). For
example, if we keep only the dominant SUSY mass corrections and
simplify the calculation by assuming a degenerate spectrum we
find \newline

\vspace{0.5cm}

$\bf T>M_{SUSY}$
\begin{eqnarray}
B&=& (12+24\tilde{\alpha})\mu_{u_{L}}\label{eq:nosphal3}\\ L&=&
(261+234\tilde{\alpha}+\frac{27}{1+2\tilde{\alpha}})\frac{\mu_
{u_{L}}}{4} \label{eq:nosphal4}
\end{eqnarray}
where $\tilde{\alpha}$ is the common factor taking account of the
supersymmetric mass effects. It is straightforward to use
eqs(\ref{eq:susychem}) to calculate the values of B and L for any
supersymmetric spectrum.
\subsubsection{Sphalerons}
Now let us consider the realistic case that the Sphaleron and
$\delb$  interactions are in thermal equilibrium for any $T\gsim
\mw$. If one or more of the operators (\ref{eq:rpbary}) are in
thermal equilibrium together with the Sphaleron and Higgs
interactions of the MSSM at any $T\gsim M_{W }$ then the
symmetries of Eq.(\ref{eq:cons3}) are reduced to the conserved
quantum numbers   \begin{equation}
L_i-L_j.
\label{eq:cons2a}
\end{equation}
Including the sphalerons gives us the additional chemical
potential equation (\ref{eq:chemsphal}).

\vspace{0.7cm}

\noindent {$\bf T\gsim T_C, T>>M_{SUSY}$}

\vspace{0.4cm}

\noindent In this limit the leptons are massless and the
supersymmetric masses may be neglected.  This leaves us with the
trivial solution  \begin{equation}
B=L=0,
\end{equation}
as well as $\mu_{uL}=\mu_0=\mu=0$. However, as in Section
\ref{sec-b-l} we do not necessarily have $\mu_i$=0. Non-zero
$\mu_i$ can regenerate a baryon-asymmetry below $T_C$ and we
discuss this below. With {\it only} a pre-existing baryon-number
asymmetry the $\mu_i=0$, and in order to avoid erasure of the
baryon and/or lepton asymmetry we must demand the strong bound
(\ref{eq:barybound3}) on the Yukawa couplings of Eq.(31).

\subsubsection{Leptogenesis, $T\lsim T_{C}$}
\label{sec-lep}
We now consider the implications of relaxing assumption (A2) of
section 4 and allow for GUT-scale leptogenesis (or family-asymmetric $B+L_i$
generation). If, as was assumed above, all
supersymmetric states are degenerate then baryon number
regeneration will occur only through the lepton mass effects
discussed above. However we expect the slepton and sneutrino
states to have mass splittings. The most significant contribution
comes from radiative corrections involving the (flavour
dependent) Yukawa couplings which are expected to give the
sleptons a much larger splitting than the leptons. We therefore
neglect $\Delta\mu$ compared to $\Delta{\tilde\mu}=\mu-
\sum{\tilde\alpha}_i\mu_i$. Allowing for a common slepton,
sneutrino flavour dependent mass matrix gives,  in  the case of
the $\delb$, \rpv\ operators in thermal equilibrium, four
equations, for the five unknowns
$\mu_{uL},\,\mu_W,\,\mu_0,\,\mu$, and $\Delta{\tilde{\mu}}$.
Solving for $B$ and $L$ in terms of $\Delta {\tilde\mu}$ we
obtain
\begin{eqnarray}
B&=& (-16+\frac{1344+1400\tilde{\alpha}} {144+270\tilde{\alpha}
+125\tilde{\alpha}{2}}) \frac{\Delta{\tilde{\mu}}}{15}\\ %
L&=&(-13+\frac{1392+1450\tilde{\alpha}} {144+270\tilde{\alpha}
+125\tilde{\alpha}{2}}) \frac{\Delta{\tilde{\mu}}}{5}
\end{eqnarray}
Again, $\Delta\tilde{\mu} \not= 0$, since $\mu_i \not= 0$ and the
baryon-number reappears {\it due} to sphaleron interactions
converting $L_i$ number excess  into $B$ excess.
Therefore, if we allow for leptogenesis, or more generally $L_i-
L_j\not=0$ as a  high-temperature initial condition, then the
bound on the $\delb$,  \rpv-couplings completely collapses.

\subsection{Baryogenesis Constraints on $\dell$ Operators}
\label{sec-bar}
Here we consider the $\delli$ operators (\ref{eq:rplep}), and
assume all $\delb$ couplings (\ref{eq:rpbary}) vanish:
$\lam''\equiv0$. As an example, we  introduce a $\Delta
L_1\not=0$ operator $[L_1Q_i{\bar D}_j]_F$. This operator is in
thermal equilibrium together with the Sphaleron configurations
and the Higgs interactions provided the relevant coupling
violates the bound (\ref{eq:barybound2}), where we must replace
the squark mass in (\ref{eq:barybound2}) with the appropriate
scalar fermion mass. In this case two of the three conserved
quantum-numbers (\ref{eq:cons3}) remain  \begin{equation} \third
B-L_2,\quad \third B-L_3,
\label{eq:cons2b}
\end{equation}
and the initial baryon-asymmetry as well as the lepton-asymmetry
in $L_2$ and $ L_3$ are not washed out. This is obvious from the
chemical potential analysis, since the addition of the
$[L_1Q_i{\bar D}_j]_F$ interaction in thermal  equilibrium
implies ($\mu_1+\mu_{dL}-\mu_{dR}=)$ $\mu_1=0$, but imposes no
condition on $\mu_{2,3}$ and hence $\mu=\mu_2+\mu_3$ may be non-zero.
Thus the result of Eqs.(\ref{eq:bsm}-\ref{eq:barylow})
remains. Notice that there are {\it eleven} operators for which
only $\Delta L_1\not =0$. Thus we can allow for  eleven operators
violating the inequality (\ref{eq:barybound2}) and still conserve
the symmetries (\ref{eq:cons2b}) and maintain the GUT-scale
baryon-asymmetry. In addition, one can also allow for the eleven
operators for which $\Delta L_2\not =0$ to violate
(\ref{eq:barybound2}) leaving one remaining symmetry
\begin{equation}
\third B-L_3,
\label{eq:cons1}
\end{equation}
which is still sufficient to maintain the GUT-scale baryon-asymmetry.
In allowing for these operators to be in thermal
equilibrium, we must at the same time require the operators
violating $L_3$ to satisfy the bound (\ref{eq:barybound2})
otherwise we would also violate the remaining symmetry
(\ref{eq:cons1}). This explicitly violates assumption (A1). In
order to have detectable effects of \rpv\ at colliders we are
interested in Yuakawa couplings $\lam\gsim10{-3}$. If for
example $\lam_{122}=\lam_{211}=10{-2}$, for which $\Delta
L_1\not=0$ and $\Delta L_2\not=0$ respectively, we must require
$\lam_{311}<5\cdot10{-7}$ and thus
\begin{equation}
\frac{\lam_{122}}{\lam_{311}}> 2\cdot 104
\label{eq:hierarchy}
\end{equation}
in direct violation of assumption (A1). However, we do not find
this hierarchy unreasonable. An obvious possibility is that there
is a  {\it flavor dependent} symmetry with  the exact
conservation of a particular family lepton-number, here giving
$\lam_{311}=0$. Even without such a symmetry it is perfectly
reasonable to expext large differences between the various lepton
couplings. In the SM we have for the Yukawa couplings of
Eq.(\ref{eq:higgs}) \begin{eqnarray}
\frac{h_{top}}{h_{elec}}&={\displaystyle{\frac{m_{top}}{m_{ele
c}}}}& \simeq 3\cdot105,\\
\frac{h_{top}}{h_{up}}&= {\displaystyle{\frac{m_{top}}{m_{up}}}}&
\simeq 3\cdot104,
\label{eq:hierarchy2}
\end{eqnarray}
where we have assumed a top quark mass of $150\,GeV$. This
hierarchy is of the same order as that of the \rpv, $\delli$
couplings in  Eq.(\ref{eq:hierarchy}).  We note also that the
magnitude of $h_{elec}$ is itself $\approx 5\cdot 10{-6}$, quite
close to the bound.

Thus we conclude, there is no reason  to rule out a subset of
$\Delta L_i \not= 0 $ operators on the grounds of baryogenesis.
These operators may be present at a level which would cause the
lightest supersymmetric particle to decay within the experimental
apparatus thus changing the characteristic signal for
supersymmetry \cite{rpcoll,butter,rp3}.

Although we have concentrated in this section on the question of
erasure of a pre-existing baryon number asymmetry we note that,
since the arguments for the $\Delta L_i \not= 0 $ operators do
not rely on any cancellation between the $\mu_i$, leptogensis
with late conversion to a baryon asymmetry is possible in this
cases as well and does not change any of the conclusions.

\section{SUSYGUTS}
\label{sec-susy}
Our discussion of Section \ref{sec-bar} showed that the conservation of
$\frac{B}{3}-L_{i}$ is sufficient to allow a pre-existing baryon number excess
to persist in the presence of a subset of lepton-number violating operators.
This conclusion would be fallacious if there should exist significant flavour
violating processes in the lepton sector. Such processes would also invalidate
the conclusions of Section 4.3.3. It has been pointed out \cite{slepton} that
in
many (non-minimal) SUSY-GUTs there are just such processes coming from new
lepton-number violating processes involving new super-heavy chiral
supermultiplets, X, with mass, $M_{X}$, of order the GUT scale. These
multiplets
have lepton-number flavour violating couplings which we generically denote by
k.
Provided the processes producing the original baryon number excess occur at a
temperature below the mass of these X fields it would appear that their
lepton-flavour violating interactions will not affect the conclusions of
Section
\ref{sec-bar}. This is not the case\footnote{We thank L.Hall for drawing our
attention to this point.} because the new interactions induce lepton-flavour
violating masses${2}$ in the slepton sector of $O(\frac{k M_{SUSY}}{4 \pi})2$
{\it not} suppressed by inverse powers of $M_{X}$  \cite{slepton}. In this
section we consider whether these effects are likely to vitiate the conclusions
of Section \ref{sec-bar}. The important question is whether slepton flavour
violating masses induce lepton-flavour violating processes which are in thermal
equilibrium at some time after baryogenesis. Note first that above $T_{C}$
electroweak breaking effects vanish and so the charged and neutral sleptons
mass
matrix may be simultaneously diagonalised. Thus, above $T_C$, flavour changing
effects will not occur through gauge interactions. Since we have assumed that
baryogenesis occurs after the lepton-number violation directly generated by X
exchange processes have dropped out of equilibrium we must consider
lepton-flavour violating processes involving sleptons and leptons  mediated by
Higgs scalars and Higgsinos which have mass of $O(M_{SUSY})$. Their
lepton-flavour violating couplings are $\frac{m_{l_{i}}}{v} sin(\theta )$ where
$\theta$ is the slepton or lepton  mixing angle induced by flavour violating
entries in the slepton mass matrix. The relation of the mixing angle to the
flavour violating masses depends on the splitting between slepton generations.
The normal (flavour diagonal) Higgs interactions $\propto \frac{m_{l_{i}}}{v}$
lead to mass$2$ splitting of $O( \frac {m_{l_{i}}} {v} ln( \frac {M_{X}}
{M_{SUSY}}))2$, logarithmically enhanced relative to the heavy particle
flavour
violating contribution. If $k\le \frac{m_{l_{i}}}{v}$ the expectation for
$\sin(\theta)$ is $\le O(\frac{1}{ln2(\frac{M_{X}}{M_{SUSY}}})$. If k is
larger
there is no manifest suppression although the actual value of $\theta$ will
depend on the structure of the matrix of couplings k relative to the matrix of
Yukawa couplings involving the light Higgs scalars.

The dominant process involving the Higgsino lepton-flavour
violating coupling is the 2-1 process and the condition this
should not be in equilibrium is given in eq(40)
\begin{equation}
(\frac{m_{l_{i}}}{v}sin(\theta))\le
5.10{-7}(\frac{\tilde{m}}{1TeV})
\label{eq:ratel}
\end{equation}
This condition requires $sin(\theta)\le 10{-3}$ where we have
used the dominant coupling $\propto m_{\tau}$. If this is not
satisfied the lepton-flavour violating processes descending from
the SUSY-GUT processes will be sufficient to erase any pre-existing
baryon number excess. However, the bound is relatively
mild and could be satisfied in specefic SUSY-GUTS; after all,
mixing angles of this magnitude occur in the quark sector. (In
the case of small couplings k, we saw above that small mixings
always result.) Thus we cannot draw general conclusions and must
consider specefic models. In this context we note that string
inspired unification schemes need not lead to new states X of the
type which can lead to lepton-flavour violating couplings for
they can generate a SM structure directly at the compactification
scale. Also, we note that the minimal $SU(5)$ alone does not
introduce couplings leading to  such $L_i$ violating slepton
masses.

\section{ Conclusions}
We have analysed the effect of the $B$ and $L$ violating
sphaleron interactions on the overall baryon- and lepton-number
of the universe. In the SM and MSSM we find that an early baryon-number
asymmetry survives to the present and leads to non-zero
$B$ {\it and} $L$, provided there is a component of the  baryon-asymmetry
generated in a $(B-L)\ne 0$ channel. This constraint is not necessary if the
early matter-genesis induces non-zero
$L_i-L_j$, for example in family  non-symmetric leptogenesis, for
then we obtain non-zero values for $B$ and $L$ {\it even if} $(B-
L)=0$ for all temperatures.

We have also considered the constraints on \rpv\ couplings in
supersymmetric models which follow from the requirement that a
baryon number asymmetry be present at the present era. For late
baryogenesis at the electroweak scale there is no constraint on
such couplings. The survival of an early baryogenesis does impose
conditions on $\delb$, and $\dell$, \rpv-couplings. However, the
bounds are much more model dependent than has hitherto been
discussed and are completely evaded in physically reasonable
scenarios. If an initial (flavour asymmetric) lepton number
asymmetry is generated there is {\it no} bound on the presence
of $\Delta B \ne 0$ \rpv\ operators. In the case of $\Delta L \ne
0$ operators the bounds depend sensitively on the flavour
dependence assumed for their coefficients and vanish completely
if a particular lepton flavour quantum number is conserved. As
a result such operators may be present at a level which causes
the lightest supersymmetric particle to decay within the detector
giving rise to non-standard signals for supersymmetry.

\bigskip

{\bf Note added}

\noindent After completing this work we received a new analysis
\cite{campbell2} by the authors of Ref.\cite{campbell}. in which
the full calculation of Secs 4.2.1 and 4.2.2 is also presented,
in agreement with the results presented here. However their
analysis does not take account of mass effects, or of the
possibility of early leptogenesis and lepton-flavour violating,
\rpv\ couplings and so their conclusions differ from those
presented here.

\bigskip

{\bf Acknowledgements}

\noindent We would like to thank Subir Sarkar for very helpful
and enjoyable conversations. This work was supported by the SERC.

\newpage
{\bf Table^1:} Deviation of the mass correction factor from one at $T=\mw$ and
$T=T_C$. $\phi0$ and $\phi-$ denote the neutral and charged Higgs
respectively. For the numerical evaluation we have chosen a Higgs mass of $400
\, GeV$ at $T_C$, and we have chosen
$m_{\phi0}=100\,GeV$ and $m_{\phi-}=\mw$ at $\mw$.

\bigskip

{\bf  Table^2:} Chemical potential notation for the particles of
the Standard Model.

\bigskip

{\bf  Fig.^1:} $2\ra2$ baryon- or lepton-number violating
interactions for the operators (a) $[{\bar U}{\bar D}{\bar
D}]_F$, (b) $[LQ{\bar D}]_F$, and (c) $[LL{\bar E}]_F$. We have
used the tilde notation to denote supersymmetric particles and
have omitted generation indices. The circled vertices are those
due to \rpv.

\bigskip

{\bf  Fig.^2:} As in Fig.^1, except for the $2\ra1$ process.
\vspace{3cm}

\bigskip

\begin{tabular}{|c|ccccc|} \hline
  & $e$ & $\mu$ & $\tau$ & $d$  & $s$   \\ \hline
$1-\alpha(m_k/M_W)$ & $2.0\cdot10{-11}$&$2.6\cdot 10{-7}$& $
7.4\cdot10{-5}$& $ 2.1\cdot10{-10}$&$ 2.1\cdot10{-6}$   \\ $1-
\alpha(m_k/T_C)$ &0&0&0 & 0&0\\
\hline
\end{tabular}
\smallskip
\begin{tabular}{|c|ccccccc|} \hline
& $b$   &  $u$ & $c$ &  $t$ & W& $\phi0$ & $\phi-$\\ \hline $1-
\alpha(m_k/M_W)$ &
$4.6\cdot10{-4}$ &
$6.2\cdot10{-10}$&$ 5.2\cdot10{-5}$ &0.38& 0.40 &0.48& 0.40 \\
$1-\alpha(m_k/T_C)$ &0&0&0&0 &0 & 0.58& 0.58\\
\hline
\end{tabular}
\vspace{1cm}
\indent \hspace{6cm} {\bf Table 1}
\vspace{3.0cm}
\begin{tabular}{|c|cccccccccc|} \hline
& $e_{Li}$ & $e_{Ri}$ & $\nu_{Li}$ & $u_L$ & $u_R$ & $d_L$ &
$d_R$  & $W+$ & $\phi0$ & $\phi-$\\ \hline
Chem. Pot. & $\mu_{Li} $ & $\mu_{Ri} $& $\mu_i $&$\mu_{uL} $
&$\mu_{uR} $ &  $\mu_{dL} $& $\mu_{dR} $& $\mu_W $& $\mu_0 $& $\mu_-
 $ \\ \hline
\end{tabular}
\vspace{1cm}
\indent \hspace{6cm} {\bf Table 2}
\end{document}